\documentclass[preprint,aps,superscriptaddress]{revtex4}
\pdfoutput=1
\usepackage{amssymb}
\usepackage{amsmath}
\usepackage{graphicx}
\usepackage{color}
\usepackage{subfigure}

\newcommand\bea{\begin{equation}}
\newcommand\eea{\end{equation}}
\newcommand\beq{ \begin{eqnarray} }
\newcommand\eeq{ \end{eqnarray} }

\input{tcilatex}

\begin{document}

\title{$d+id$ Holographic Superconductors}
\author{Jiunn-Wei Chen\thanks{%
E-mail: jwc@phys.ntu.edu.tw}}
\affiliation{Department of Physics and Center for Theoretical Sciences, National Taiwan
University, Taipei 10617, Taiwan}
\author{Yu-Sheng Liu\thanks{%
E-mail: mestelqure@gmail.com}}
\affiliation{Department of Physics and Center for Theoretical Sciences, National Taiwan
University, Taipei 10617, Taiwan}
\author{Debaprasad Maity\thanks{%
E-mail: debu.imsc@gmail.com}}
\affiliation{Department of Physics and Center for Theoretical Sciences, National Taiwan
University, Taipei 10617, Taiwan}
\affiliation{Leung Center for Cosmology and Particle Astrophysics\\
National Taiwan University, Taipei 106, Taiwan}

\begin{abstract}
A holographic model of $d+id$ superconductors based on the action proposed
by Benini, Herzog, and Yarom [arXiv:1006.0731] is studied. This model has a
charged spin two field in an AdS black hole spacetime. Working in the probe
limit, the normalizable solution of the spin two field in the bulk gives
rise to a $d+id$ superconducting order parameter at the boundary of the AdS.
We calculate the fermion spectral function in this\ superconducting
background and confirm the existence of fermi arcs for non-vanishing
Majorana couplings. By changing the relative strength $\gamma $ of the $d$
and $id$ condensations, the position and the size of the fermi arcs are
changed. When $\gamma =1$, the spectrum becomes isotropic and the spectral
function is s-wave like. By changing the fermion mass, the fermi momentum is
changed. We also calculate the conductivity for these holographic $d+id$
superconductors where time reversal symmetry has been broken spontaneously.
A non-vanishing Hall conductivity is obtained even without an external
magnetic field.
\end{abstract}

\maketitle

\begin{flushright}
\end{flushright}

%\date{\today}

\section{Introduction}

The mechanism of the high temperature superconductors (SC's) is an unsolved
mystery in physics \cite{Lee:2006}. High temperature SC's are layered
compounds with copper-oxygen planes and are doped Mott insulators with
strong electronic correlations. The pairing symmetry is unconventional and
there is a strong experimental evidence showing that it is largely d-wave 
\cite{Tsuei:2000}. It is speculated that the pairing between electrons is
mediated via strong anti-ferromagnetic spin fluctuations in the system. But
the problem is difficult due to the strong-coupling nature of the system.
Although significant progress has been made in the last many years \cite%
{Anderson,gaugetheories}, alternative approaches may be valuable to tackle
the problem.

An interesting alternative approach is the holographic correspondence
between a gravitational theory and a quantum field theory, which first
emerged under the framework of AdS/CFT correspondence \cite%
{Maldacena:1997re,Gubser:1998bc,Witten:1998qj}. This method has provided us
a useful and complimentary framework to describe strong interacting systems
without the sign problem (see e.g. \cite%
{Policastro:2001yc,Herzog:2006gh,Liu:2006ug,Gubser:2006bz,Herzog:2007ij,Hartnoll:2007ih,Hartnoll:2007ip,Hartnoll:2008hs}%
). In the original top-down approach within the AdS/CFT framework, both the
gravity side and the field theory side were precisely know in string theory
and gives us much deeper insight on this correspondence. But later in
bottom-up approach we assume that the correspondence exists among the
different pair of theories and try to make predictions from one side of the
correspondence.

Recently, a gravitational model of hairy black holes \cite%
{Gubser:2005ih,Gubser:2008px} has been used to model s-wave high temperature
SC's \cite{Hartnoll:2008vx,Hartnoll:2008kx,Horowitz:2009ij,Gubser:2008zu}.
In those class of models the Abelian symmetry of a complex scalar field is
spontaneously broken (i.e. the Higgs mechanism) below some critical
temperature. The Meissner effect was soon observed by including a magnetic
field in the background \cite{Nakano:2008xc,Albash:2008eh}. A similar
construction of holographic multiband superconductor has also been
considered recently in \cite{debumulti}. The multiband is coming from the
condensation of a fundamental scalar field multiplet under non-abelian gauge
group. The effect of the superconducting condensate on the holographic fermi
surface has further been studied by calculating fermionic spectral functions 
\cite{Chen:2009pt,Faulkner:2009am,Gubser:2009dt}. Interestingly, the
properties of spectral function appeared to have similar behavior to that
found in the angle resolved photo-emission experiment. Analogously,
holographic p-wave superconductors have also been proposed by coupling a $%
SU(2)$ Yang-Mills field to the black hole background, where a vector hair
develops in the superconducting phase \cite%
{Gubser:2008wv,Roberts:2008ns,Martin:2009plb,Pallab}. Properties of
fermionic spectral function has also been studied in p-wave superconducting
background \cite{fermionp}. Also, there are attempts to build holographic
d-wave SC's by spontaneously breaking the Abelian symmetry of a charged spin
two field \cite{Chen:2010mk,herzog,Benini:2010qc,Benini:2010pr} and study
different properties of the systems \cite{Zeng:2010fx,Zeng:2010vp,Pan:2011ns}%
. In addition to the bottom up approach mentioned above, there are various
top-down constructions of those condensed matter like systems by considering
D-brane configurations in the AdS black hole background in the string theory
framework \cite{Martin2}.

In this work we will be interested in studying more detailed properties of
holographic d-wave superconductors using the action of \cite%
{Benini:2010qc,Benini:2010pr}. The most challenging problem to construct the
holographic model of d-wave SC's is that the consistent action for the
charged, massive spin two field in a general curve background is still not
known. Although the action used in \cite{Benini:2010qc,Benini:2010pr} has
some interesting features, such as having the right degrees of freedom and
being ghost-free, it could be unstable or acausal for general gauge field
configurations. It is argued that these problems may be cured by adding
higher dimensional operators \cite{Benini:2010pr}.

In this work, we will study properties of the holographic $d+id$ SC's based
on the action of \cite{Benini:2010qc,Benini:2010pr}. This is motivated by
the observations of spontaneous breaking of time reversal invariance in, for
example, the YBCO high temperature superconductor \cite%
{one,two,three,four,five}. The time reversal breaking is thought to occur
due to a complex combination of the d-wave condensates: $%
d_{x^{2}-y^{2}}+id_{xy}$. One of the possible interesting consequences of
this complex d-wave condensation is that the system has a Hall conductivity
even in the absence of an external magnetic field \cite{tone, ttwo}. By
using the holographic framework, in this note we are interested in
calculating the fermion spectral function, and the normal and Hall
conductivity in the holographic $d+id$ SC's.

Our paper is organized as follow: In section \ref{sec2} we will first
introduce our model following \cite{Benini:2010pr}. We consider the
effective action for the charged spin two field in a four dimensional
AdS-Schwarzschild black hole. In order for the system to be ghost free, we
will consider the probe limit. Subsequently in section \ref{sec3} we will
obtain the normalizable background spin two field configuration which leads
to a dual $d+id$ superconducting phase of a strongly coupled system living
on the boundary of AdS. Once we identify our required dual complex d-wave
condensate background, we can study various linear transport properties of
the system by considering linear order fermionic and gauge field
perturbations. In section \ref{sec4} we will study the fermionic spectral
function and discuss various properties depending upon various
parameters such as the Majorana coupling, mass of the fermion etc. In
section \ref{sec5} we will study the conductivity of the superconducting
background. As we have mentioned earlier, because of spontaneous violation
of time reversal symmetry due to the background condensation, we found a
non-vanishing optical Hall conductivity. In the final section we will
conclude with some remarks and discuss future directions.

\section{The Model}

\label{sec2} As we have mentioned in the introduction, the model that we are
going to present is in the similar spirit of the holographic dual
construction of an s-wave superconductor in an AdS black hole spacetime. It
is well known that the phases of d-wave superconductivity can be described
by a low energy effective theory of charged spin two tensor field in the
framework of Landau-Ginzburg theory. At low temperature, the background
solution of this charged spin two field yields the d-wave order parameter
for the high temperature superconductivity. Analogously one particular
construction of a holographic model of d-wave superconductor has been
considered first in \cite{Chen:2010mk,herzog} by taking a charged symmetric
traceless spin-2 field in an AdS-black hole background. Subsequently, in
spite of having long-standing technical and also conceptual problems, a more
refined form of the charge spin-2 tensor field effective action has been
proposed and studied in the context of pure d-wave superconductor \cite%
{Benini:2010pr}. Our main goal in this report is to study the properties of
the more general $d+id$ superconducting background in AdS/CFT framework. One
of the theoretical motivations to consider these kinds of systems is to
discover some universal low energy properties of general strongly coupled
field theories. Keeping this motivation in mind, we start by considering the
action in the bulk containing the gravity part and the matter part as 
\begin{equation}
S=\frac{1}{2\kappa ^{2}}\int d^{4}x\sqrt{-g}\left\{ \left( R+\frac{6}{L^{2}}%
\right) +\mathcal{L}_{m}\right\} ,  \label{Lagrangian}
\end{equation}%
where $R$ is the Ricci scalar, the $6/L^{2}$ term gives a negative
cosmological constant and $L$ is the AdS radius which will be set to unity
in the units that we use. $\kappa ^{2}=8\pi G_{N}$ is the gravitational
coupling. The Lagrangian for the matter fields includes a spin two field and
a spin half fermion field. Both of them are charged under U(1) gauge field
and can be massive. We write 
\begin{equation}
\mathcal{L}_{m}=\mathcal{L}_{b}+\mathcal{L}_{f}\ .
\end{equation}

The consistent construction of a charged spin two field theory in curved
spacetime background is a long-standing problem. There have been lot of
studies in constructing consistent interacting higher spin field theories 
\cite{higherspin}. As we have mentioned, the authors in Ref.\cite%
{Benini:2010pr} have proposed a unique Lagrangian for a charged spin-2 field
in AdS space with the motivation in studying its dual field theoretic
properties. In this report we will adopt their construction with the
Lagrangian

\begin{equation}
\begin{split}
\mathcal{L}_{b}& =-|D_{\rho }\varphi _{\mu \nu }|^{2}+2|D_{\mu }\varphi
^{\mu \nu }|^{2}+|D_{\mu }\varphi |^{2}-\big[D_{\mu }\varphi ^{\ast \mu \nu
}D_{\nu }\varphi +\text{c.c.}\big] \\
& \quad -m^{2}\big(|\varphi _{\mu \nu }|^{2}-|\varphi |^{2}\big)+2R_{\mu \nu
\rho \lambda }\varphi ^{\ast \mu \rho }\varphi ^{\nu \lambda }-R_{\mu \nu
}\varphi ^{\ast \mu \lambda }\varphi _{\lambda }^{\nu } \\
& \quad -\frac{1}{d+1}R|\varphi |^{2}-iqF_{\mu \nu }\varphi ^{\ast \mu
\lambda }\varphi _{\lambda }^{\nu }-\frac{1}{4}F_{\mu \nu }F^{\mu \nu },
\end{split}
\label{L}
\end{equation}%
where $\varphi _{\mu \nu }$ is the spin two field with $\varphi _{\mu \nu
}=\varphi _{\nu \mu }$, $\varphi =\varphi _{\;\;\mu }^{\mu }$, $D_{\mu }$ is
the covariant derivative ($D_{\alpha }\varphi _{\mu \nu }=\left( \partial
_{\alpha }+iqA_{\alpha }\right) \varphi _{\mu \nu }$ in flat space), $d(=3)$%
is the spatial dimension of the bulk and $F_{\mu \nu }$ is the field
strength tensor of the gauge field. As it has been discussed in detail in 
\cite{Benini:2010pr}, the above action describes dynamics of the correct
number of degrees of freedom only when the background is Einstein's
manifold. So, the energy density of the higher spin field configuration
should be very small compared to the background energy density. This
essentially means we have to maintain the probe limit of spin-2 and fermion
fields in a specific gravitation background. However, there still exists
serious issues in dealing with the above action. For generic gauge field
background, even though it has correct number of dynamical degrees of
freedom, the equation motion loses its hyperbolicity or causality. The
general belief is that these violations can be ameliorated by considering
higher derivative operators in our Lagrangian. This requirement makes it
very complicated to construct a fully consistent Lagrangian for the higher
spin field in a curved background. However as has been argued in \cite%
{Benini:2010pr}, we will adopt the effective field theory point of view,
where all these effects could be very small in the limit of very low gauge
field strength. This is essentially the limit in which we will be
considering in our subsequent discussions.

In addition to above Lagrangian for the spin-2 field, we also consider the
spin 1/2 fermion Lagrangian as

\begin{equation}
\mathcal{L}_{f}=i\overline{\Psi }\left( \Gamma ^{\mu }D_{\mu }-m_{\zeta
}\right) \Psi +\eta ^{\ast }\varphi _{\mu \nu }^{\ast }\overline{\Psi ^{c}}%
\Gamma ^{\mu }D^{\nu }\Psi -\eta \overline{\Psi }\Gamma ^{\mu }D^{\nu
}\left( \varphi _{\mu \nu }\Psi ^{c}\right) .  \label{Lf}
\end{equation}%
The bulk Gamma matrices $\Gamma ^{\mu }$ satisfies the Clifford algebra $%
\{\Gamma ^{\mu },\Gamma ^{\nu }\}=2g^{\mu \nu }$. The $U(1)$ gauge symmetry
demands that the charges of the fermion and spin two field are related by $%
2q_{\zeta }=q$. It is know that $\eta $ dependent Majorana coupling helps
the fermion spectral function to develop a gap. It has been argued that
although there are more terms of the same dimension as these $\eta $ terms,
only $\eta $ dependent terms gives rise to an anisotropic gap. Thus, we have
dropped the other terms for simplicity.

Under the field redefinition 
\begin{eqnarray}
A_{\mu } &\rightarrow &A_{\mu }/q,\ \varphi _{\mu \nu }\rightarrow \varphi
_{\mu \nu }/q,  \notag \\
\Psi &\rightarrow &\Psi /q,\ \eta \rightarrow q\eta ,  \label{scaling}
\end{eqnarray}%
$\mathcal{L}_{m}$ can then be written as%
\begin{equation}
\mathcal{L}_{m}\rightarrow \mathcal{L}_{m}/q^{2}.
\end{equation}%
Thus, as $q\rightarrow \infty $ while the $\mathcal{L}_{m}$ remains finite,
we can work in the so called probe limit to ignore the back reaction of the
the matter fields $A_{\mu }$, $\varphi _{\mu \nu }$, and $\Psi $. Also, in
the probe limit, the $q$ dependence of observables can be recovered through
the above simple scaling. Thus, we will work in the probe limit from now on.
In this limit, the gravitational field equation satisfies the Einstein
equation 
\begin{equation}
R_{\mu \nu }=\frac{2\Lambda }{d-1}\,g_{\mu \nu }\;,  \label{EinsteinEqn}
\end{equation}%
where $\Lambda =-3/L^{2}$. This yields the AdS Schwarzschild black hole
solution with the metric 
\begin{equation}
ds^{2}=\frac{L^{2}}{z^{2}}\left( -f(z)dt^{2}+d\vec{x}_{d-1}^{2}+\frac{dz^{2}%
}{f(z)}\right) ,  \label{metric}
\end{equation}%
with%
\begin{equation}
f(z)=1-\left( \frac{z}{z_{h}}\right) ^{d},
\end{equation}%
where $z=z_{h}$ is the black hole horizon and $z=0$ is the boundary. The
Hawking temperature can be expressed as 
\begin{equation}
T=\frac{d}{4\pi z_{h}}.
\end{equation}%
An important point to note is that, the action is invariant under the scaling
transformations%
\begin{eqnarray}
(t,x,y,z,T) &\rightarrow &(ct,cx,cy,cz,T/c),  \notag \\
\left( A_{\mu },\varphi _{\mu \nu },\Gamma ^{\mu },\Psi \right) &\rightarrow
&\left( A_{\mu }/c,\varphi _{\mu \nu }/c^{2},c\Gamma ^{\mu },\Psi \right) ,
\label{rescaling}
\end{eqnarray}%
which determine the conformal dimension of each field. In the following
section, we will consider this black hole background in order to solve the
equations of motion for the spin-2 and electromagnetic gauge field in the
probe limit.

\section{$d+id$ Condensate}

\label{sec3} As we know from the standard AdS/CFT dictionary, the bulk
massive spin-2 field corresponds to a spin-2 operator in the boundary field
theory with fixed conformal dimension. The conformal dimension is fixed by
the spin-2 mass. This boundary spin-2 dual operator under boundary Lorentz
transformations is to be identified with the d-wave order parameter. Our
main interest here is to describe the d-wave SC in the dual boundary field
theory. Now since the spacetime background that we have considered has a
translation symmetry in the boundary direction, the condensation on the $x$-$%
y$ plane on the boundary should also have translational invariance. However
rotational symmetry should be spontaneously broken down to $Z(2)$ with the
d-wave like condensate changing its sign under a $\pi /2$ rotation on the $x$%
-$y$ plane. To incorporate these features, we use an ansatz for the
symmetric traceless $\varphi _{\mu \nu }$:%
\begin{eqnarray}
\varphi _{xx} &=&-\varphi _{yy}=\frac{1}{2z^{2}}\,\psi _{1}(z),  \notag \\
\varphi _{xy} &=&\varphi _{yx}=\frac{1}{2z^{2}}\,\psi _{2}(z),
\end{eqnarray}%
and $\varphi _{\mu \nu }=0$ for $\mu $,$\nu \neq x,y$, and for the gauge
field $A_{\mu }$: 
\begin{equation}
A=A_{\mu }\,dx^{\mu }\equiv \phi (z)\,dt\,.  \label{ansatz}
\end{equation}%
Under a $\theta $ angle rotation in the $x$-$y$ plan, 
\begin{equation}
\left( 
\begin{array}{c}
\varphi _{xx} \\ 
\varphi _{xy}%
\end{array}%
\right) \rightarrow \left( 
\begin{array}{cc}
\cos 2\theta & -\sin 2\theta \\ 
\sin 2\theta & \cos 2\theta%
\end{array}%
\right) \left( 
\begin{array}{c}
\varphi _{xx} \\ 
\varphi _{xy}%
\end{array}%
\right) .  \label{rotation}
\end{equation}%
If $\varphi _{xy}/\varphi _{xx}$ is real, then we can always rotate the
coordinates in the $x$-$y$ plan such that $\varphi _{xy}=0$ and then make $%
\varphi _{xx}$ real by a gauge transformation. Otherwise $\varphi _{xx}$ and 
$\varphi _{xy}$ will both exist.\ The equations of motion of the gauge field
and the spin two field are %\label{EOMofL}
\begin{subequations}
\begin{equation}
\phi ^{\prime \prime }+\frac{3-d}{z}\phi ^{\prime }-\frac{q^{2}}{z^{2}f}%
\left( \left\vert \psi _{1}\right\vert ^{2}+\left\vert \psi _{2}\right\vert
^{2}\right) \phi =0,  \label{EOMofGauge}
\end{equation}%
\begin{equation}
\psi _{i}^{\prime \prime }+(\frac{f^{\prime }}{f}-\frac{d-1}{z})\psi
_{i}^{\prime }+(\frac{q^{2}\phi ^{2}}{f^{2}}-\frac{m^{2}}{z^{2}f})\psi
_{i}=0,\ i=1,2.  \label{EOMofSpin2}
\end{equation}%
The coupling to the fermion field through $\delta \mathcal{L}_{f}/\delta
\varphi _{\mu \nu }^{\ast }$ is dropped because there is no fermion
condensate.

Near the horizon ($\delta z=z-z_{h}\rightarrow 0$), $f(z)\simeq -d\delta
z/z_{h}$, and %\label{EOMofL}
\end{subequations}
\begin{subequations}
\begin{eqnarray}
\phi ^{\prime \prime }+\frac{q^{2}}{3z_{h}\delta z}\left( \left\vert \psi
_{1}\right\vert ^{2}+\left\vert \psi _{2}\right\vert ^{2}\right) \phi
&\simeq &0,  \notag \\
\psi _{i}^{\prime \prime }+\frac{1}{\delta z}\psi _{i}^{\prime }+\frac{%
z_{h}^{2}q^{2}\phi ^{2}}{9\delta z^{2}}\psi _{i} &\simeq &0,\ 
\end{eqnarray}%
which yields 
\end{subequations}
\begin{eqnarray}
\phi &\simeq &\frac{c_{1}}{z_{h}^{2}}\delta z,  \notag \\
\psi _{i} &\simeq &c_{2,i}+c_{3,i}\ln \delta z.  \label{x}
\end{eqnarray}%
The finiteness of $\psi _{i}$ at the horizon sets $c_{3,i}=0$.

Near the boundary $z\rightarrow 0$, $f\rightarrow 0$, and %\label{EOMofL}
\begin{subequations}
\begin{eqnarray}
\phi ^{\prime \prime }-\frac{q^{2}}{z^{2}}\left( \left\vert \psi
_{1}\right\vert ^{2}+\left\vert \psi _{2}\right\vert ^{2}\right) \phi
&\simeq &0,  \notag \\
\psi _{i}^{\prime \prime }-\frac{2}{z}\psi _{i}^{\prime }-\frac{m^{2}}{z^{2}}%
\psi _{i} &\simeq &0.
\end{eqnarray}%
$\psi _{i}$ has the asymptotic solution with two terms: 
\end{subequations}
\begin{equation}
\psi _{i}\simeq c_{4,i}z^{\Delta _{+}}+c_{5,i}z^{\Delta _{-}},\ \ \Delta
_{\pm }=\frac{3\pm \sqrt{9+4m^{2}}}{2}.  \label{cond}
\end{equation}%
One of the terms can be identified as the source and the other term
identified as the corresponding condensate. In other words, the asymptotic
behavior of the spin two field near the boundary is 
\begin{equation}
\psi _{i}(z)=z^{d-\Delta }\left[ \psi _{i}^{(s)}+O(z)\right] +z^{\Delta }%
\left[ \frac{\langle \mathcal{O}_{i}\rangle }{2\Delta -d}+O(z)\right]
\end{equation}%
where $\psi _{i}^{(s)}$ is the source, $\langle \mathcal{O}_{i}\rangle $ is
the condensate, and $\mathcal{O}_{i}$ is the field theory operator that $%
\psi _{i}$ couples to at the boundary. $m^{2}=\Delta (\Delta -d)$ and $%
\Delta $ is conformal dimension of $\langle \mathcal{O}_{i}\rangle $
(because the conformal dimension for $\psi _{i}$ is zero).

Depending on the size of $m^{2}$, there could be different scenarios:

(a) If $m^{2}>0$, $\Delta _{-}<0$, we need to set $c_{5,i}=0$ (the
sourceless condition) to keep $\psi _{i}$ finite at the boundary. Thus, $%
\phi $ has the asymptotic solution 
\begin{equation}
\phi \simeq \mu +\rho z+O(z^{2}),  \label{phi}
\end{equation}%
where the physical meaning of $\mu $ is the chemical potential while $\rho $
is the corresponding charge density. We can always use the scaling of Eq. (%
\ref{rescaling}) to set $\mu =1$. In this case $\psi _{2}(z)$/$\psi _{1}(z)$
is a $z$ independent constant. Because the combination $c_{4,2}\psi
_{1}-c_{4,1}\psi _{2}$ vanishes near the boundary. It continues to vanish
for all $z$ by the equations of motion.

(b) If $0>m^{2}>-9/4$, $\Delta _{\pm }>0$, one can either chose the $%
z^{\Delta _{+}}$ or $z^{\Delta -}$ term to be the condensate. If we choose
the $z^{\Delta _{+}}$ term to be the condensate for $\psi _{1}$, and choose
the $z^{\Delta _{-}}$ term to be the condensate for $\psi _{2}$, then Eq.(%
\ref{phi}) is still true. But then $\psi _{2}(z)$/$\psi _{1}(z)$ is no
longer a constant.

(c) If $m^{2}=0$, $\Delta _{-}=0$, if we choose the $z^{\Delta _{+}}$ term
to be the source, then Eq.(\ref{phi}) is no longer true. Hence, the physical
meaning in this case is unclear. The other choice of the source goes back to
case (a).

In case (b), the two condensates have different dimensions at the boundary.
This might imply competition between the two order parameters in the system.
However, the scaling dimension of any primary operator of a conformal field
theory has a unitarity bound \cite{minwalla}. As has been argued in \cite%
{Benini:2010pr}, this unitarity bound constrains mass of a spin-2 field as $%
m^{2}\geq 0$. Thus, in this report, we will just focus on case (a), where 
\begin{gather}
\psi _{1}(z)=iQ_{1}\psi (z),\ \psi _{2}(z)=Q_{2}\psi (z),  \notag \\
\left\vert Q_{1}\right\vert ^{2}+\left\vert Q_{2}\right\vert ^{2}=1,\psi
(z)=\psi ^{\ast }(z).
\end{gather}%
.

\begin{figure}[tbp]
\begin{center}
\includegraphics[scale=1.5]{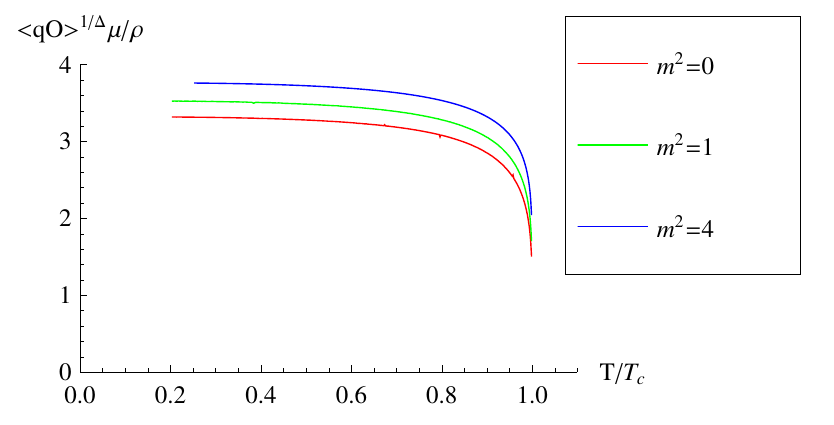}
\end{center}
\caption{$\langle q\mathcal{O}\rangle ^{1/\Delta }\protect\mu /\protect\rho $
vs. $T/T_{c}$ is shown. The combination in the $y$-axis is independent of $q$
and $\protect\mu $.}
\label{fig:condensate}
\end{figure}

We have solved the equations of motion for the spin-2 field $\psi $ and
gauge field $\phi $ using standard shooting algorithm by demanding $\mu =1$
and the normalizability of $\psi $ near the asymptotic boundary. We found
the critical temperature $T_{c}=\frac{d}{4\pi z_{hc}}$ for a fixed chemical
potential and charge, below which non-trivial bulk profile for the spin-2
field component exists. According to AdS/CFT correspondence, the d-wave
condensation ${\langle \mathcal{O}\rangle }$ in the dual boundary field
theory is identified with the coefficient of the normalizable solution of $%
\psi $. In Fig.\ref{fig:condensate} we show the condensate field ($\langle 
\mathcal{O}\rangle =\sqrt{\left\vert \langle \mathcal{O}_{1}\rangle
\right\vert ^{2}+\left\vert \langle \mathcal{O}_{2}\rangle \right\vert ^{2}}$%
) disappears above the critical temperature $T_{c}$. Once we know the
background condensation, it would be interesting to see the fermionic
response function which essentially captures the properties of the
background condensation. On the other hand, the electromagnetic perturbation
is essential to study the conductivity of this background. In what follows
we will first study the fermionic spectral function and discussed about the
gap structure of the underlying strongly coupled system depending upon the
type of fermionic coupling with a background we choose. Using the scaling
relations in Eqs.(\ref{scaling}) and (\ref{rescaling}), one can show that
the combination $\langle q\mathcal{O}\rangle ^{1/\Delta }\mu /\rho $ is
independent of $q$ and $\mu $.

\bigskip

\section{Fermion Spectral Function}

\label{sec4}

The equation of motion for the bulk fermion field looks like 
\begin{equation}
0=\big(\Gamma ^{\mu }D_{\mu }-m_{\zeta }\big)\Psi +2i\eta \varphi _{\mu \nu
}\Gamma ^{\mu }D^{\nu }\Psi ^{c}+i\eta \varphi _{\mu }\Gamma ^{\mu }\Psi
^{c},  \label{DiracEqn}
\end{equation}%
where $\varphi _{\mu }=D^{\nu }\varphi _{\nu \mu }$ and the covariant
derivative on the spinor field is

\begin{equation}
D_{\mu }\Psi =\left( \partial _{\mu }+\frac{1}{4}\omega _{\mu ,\underline{%
\lambda \sigma }}\Gamma ^{\underline{\lambda \sigma }}-iq_{\zeta }A_{\mu
}\right) \Psi
\end{equation}%
with $\omega $ the spin connection, $\Gamma ^{\mu \nu }\equiv \Gamma
^{\lbrack \mu }\Gamma ^{\nu ]}$, and with the vielbein indices underlined.
We have chosen the bulk Gamma matrices $\Gamma ^{\mu }$ representation

\begin{equation}
\begin{split}
\Gamma ^{{\underline{t}}}& =\left( 
\begin{array}{cc}
-i\sigma _{2} & 0 \\ 
0 & i\sigma _{2}%
\end{array}%
\right) ,\Gamma ^{{\underline{z}}}=\left( 
\begin{array}{cc}
\sigma _{3} & 0 \\ 
0 & \sigma _{3}%
\end{array}%
\right) , \\
\Gamma ^{{\underline{x}}}& =\left( 
\begin{array}{cc}
\sigma _{1} & 0 \\ 
0 & \sigma _{1}%
\end{array}%
\right) ,\Gamma ^{{\underline{y}}}=\left( 
\begin{array}{cc}
0 & -i\sigma _{2} \\ 
i\sigma _{2} & 0%
\end{array}%
\right)
\end{split}
\label{gamma}
\end{equation}

It is convenient to rescale $\zeta =(-g\cdot g^{zz})^{1/4}\Psi $ which
removes the spin connection completely, and to work in the momentum space
with the Fourier components:%
\begin{equation}
\zeta =e^{-i\omega t+i\vec{k}\cdot \vec{x}}\,\zeta ^{(\omega ,\vec{k}%
)}(z)+e^{i\omega t-i\vec{k}\cdot \vec{x}}\,\zeta ^{(-\omega ,-\vec{k})}(z).
\end{equation}%
We then decompose the four-component spinor to two two-component spinors: $%
\zeta =(\zeta _{1},\zeta _{2})$, such that Eq. (\ref{DiracEqn}) can be
rewritten as

\begin{equation}
\begin{split}
& D_{(1)}\zeta _{1}^{(\omega ,k_{x})}+2\eta (g^{xx})^{3/2}k_{x}\left[
\varphi _{xx-yy}\sigma _{1}\zeta _{1}^{(-\omega ,-k_{x})\ast }-i\varphi
_{xy}\sigma _{2}\zeta _{2}^{(-\omega ,-k_{x})\ast }\right] =0 \\
& D_{(2)}\zeta _{2}^{(\omega ,k_{x})}+2\eta (g^{xx})^{3/2}k_{x}\left[
\varphi _{xx-yy}\sigma _{1}\zeta _{2}^{(-\omega ,-k_{x})\ast }+i\varphi
_{xy}\sigma _{2}\zeta _{1}^{(-\omega ,-k_{x})\ast }\right] =0
\end{split}
\label{EOMofFermion}
\end{equation}%
where%
\begin{equation}
D_{(\alpha )}=\sqrt{g^{zz}}\sigma _{3}\partial _{z}-m_{\zeta }+(-1)^{\alpha
}(\omega +q_{\zeta }A_{t})\sqrt{-g^{tt}}\sigma _{2}+ik_{x}\sqrt{g^{xx}}%
\sigma _{1}.
\end{equation}%
Without losing generality, we set $k_{y}=0$ while the condensate can have
arbitrary orientations.

The spectral function is defined by the imaginary part of the trace of the
retarded green's function 
\begin{equation}
A(\omega ,\mathbf{k})\equiv \mathrm{Im}[\mathrm{Tr}(G_{R})].
\end{equation}%
To compute $A(\omega ,\mathbf{k})$, we follow the method developed in \cite%
{Liu:2009dm,Iqbal:2009fd,zaanen}and applied in \cite{Faulkner:2009am} which
leads to the simple \textquotedblleft flow equation\textquotedblright :

\begin{equation}
\begin{split}
& \sqrt{g^{zz}}\partial _{z}(iG) \\
& =2m_{\zeta }(iG)+\left[ -ik_{x}\sqrt{g^{xx}}\mathcal{A}+i(\omega \mathcal{B%
}+q_{\zeta }A_{t}\mathcal{C})\sqrt{-g^{tt}}+2k_{x}\eta (g^{xx})^{3/2}%
\mathcal{D}\right] \\
& \qquad -(iG)\left[ ik_{x}\sqrt{g^{xx}}\mathcal{A}+i(\omega \mathcal{B}%
+q_{\zeta }A_{t}\mathcal{C})\sqrt{-g^{tt}}+2k_{x}\eta (g^{xx})^{3/2}\mathcal{%
E}\right] (iG),
\end{split}
\label{EvoEqn}
\end{equation}%
where

\begin{equation}
\begin{split}
& \mathcal{A}=\left( 
\begin{array}{cccc}
1 & 0 & 0 & 0 \\ 
0 & 1 & 0 & 0 \\ 
0 & 0 & -1 & 0 \\ 
0 & 0 & 0 & -1%
\end{array}%
\right) ,\mathcal{B}=\left( 
\begin{array}{cccc}
-1 & 0 & 0 & 0 \\ 
0 & 1 & 0 & 0 \\ 
0 & 0 & 1 & 0 \\ 
0 & 0 & 0 & -1%
\end{array}%
\right) , \\
& \mathcal{C}=\left( 
\begin{array}{cccc}
-1 & 0 & 0 & 0 \\ 
0 & 1 & 0 & 0 \\ 
0 & 0 & -1 & 0 \\ 
0 & 0 & 0 & 1%
\end{array}%
\right) ,\mathcal{D}=\left( 
\begin{array}{cccc}
0 & 0 & \varphi _{xx} & -\varphi _{xy} \\ 
0 & 0 & \varphi _{xy} & \varphi _{xx} \\ 
\varphi _{xx}^{\ast } & -\varphi _{xy}^{\ast } & 0 & 0 \\ 
\varphi _{xy}^{\ast } & \varphi _{xx}^{\ast } & 0 & 0%
\end{array}%
\right) ,
\end{split}
\label{Mat in EvoEqn}
\end{equation}%
and $\mathcal{E}=\mathcal{D}^{\dagger }$. The initial condition which
satisfies the ingoing wave boundary condition at the horizon of an AdS black
hole would be 
\begin{equation}
G_{R}=z^{2m_{\zeta }}\left. G\right\vert _{z\rightarrow 0}.
\end{equation}

Using the scaling relations in Eqs.(\ref{scaling}) and (\ref{rescaling}),
one can show that the spectral function $A(\omega /q\mu ,\mathbf{k}/q\mu )$
is independent of $q$ and $\mu $, provided $\eta /q$ is also $q$ independent.

\subsection{Fermi Arc}

\begin{figure}[tbp]
\begin{center}
\includegraphics[scale=0.7]{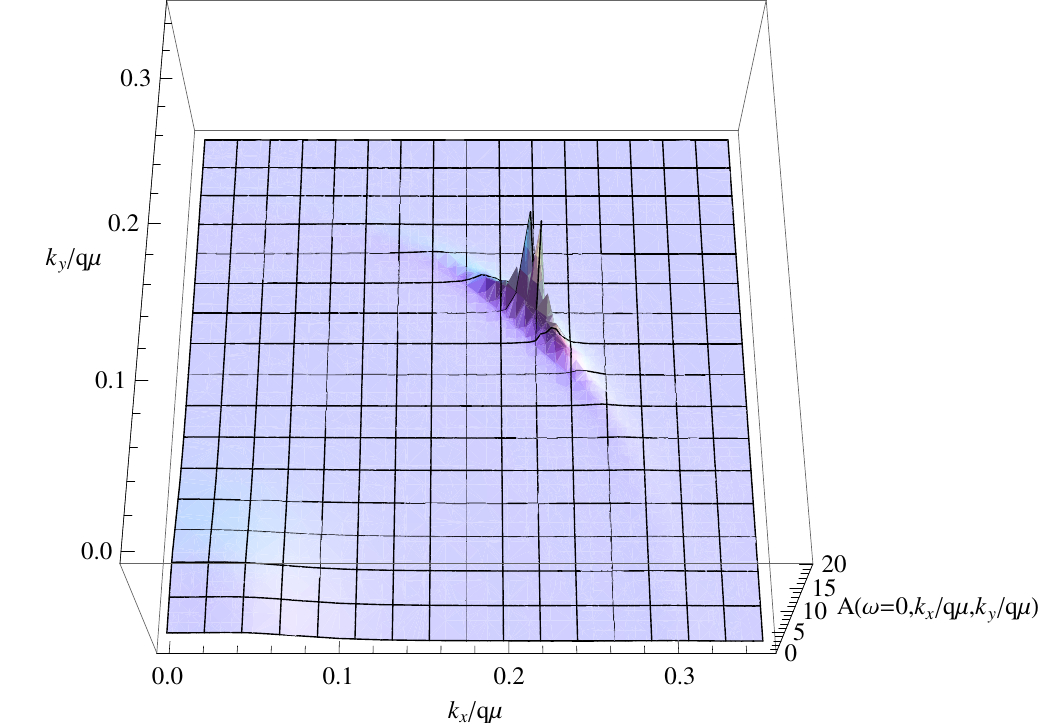}
\end{center}
\caption{The appearance of the fermi arcs in the fermion spectral density $A(%
\protect\omega =0,k_{x}/q\protect\mu ,k_{y}/q\protect\mu )$ for normal
d-wave superconductor ($\protect\varphi _{xy}=0$) with $m=0$, $m_{\protect%
\zeta }=0$, $\protect\eta =0.15q$ and $T=0.66T_{c}$. }
\label{fig:FermiArc}
\end{figure}

As mentioned above, when $\varphi _{xy}/\varphi _{xx}$ is real, we can
always rotate the coordinate in the $x$-$y$ plan such that $\varphi _{xy}=0$
and make $\varphi _{xx}$ real by a gauge transformation. The corresponding
fermion spectral function $A(\omega =0,\mathbf{k}/q\mu )$ for $\varphi
_{xy}=0$, $\eta =0.15q$, $m=0$, $m_{\zeta }=0$, and $T=0.66T_{c}$ is shown
in Fig. \ref{fig:FermiArc}.

\begin{figure}[tbp]
\begin{center}
\subfigure[$\theta=0^{\circ}$]{
\includegraphics[scale=0.75]{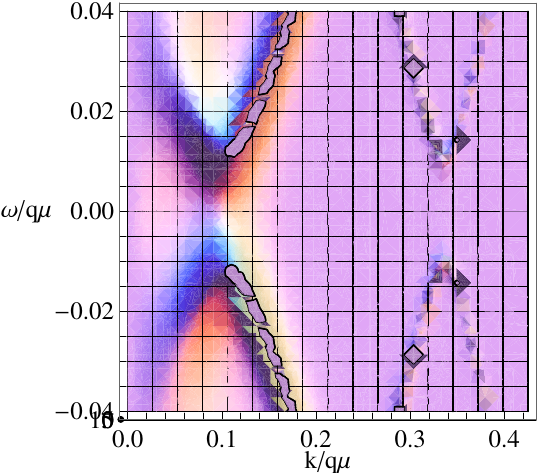}} 
\subfigure[$\theta=30^{\circ}$]{
\includegraphics[scale=0.75]{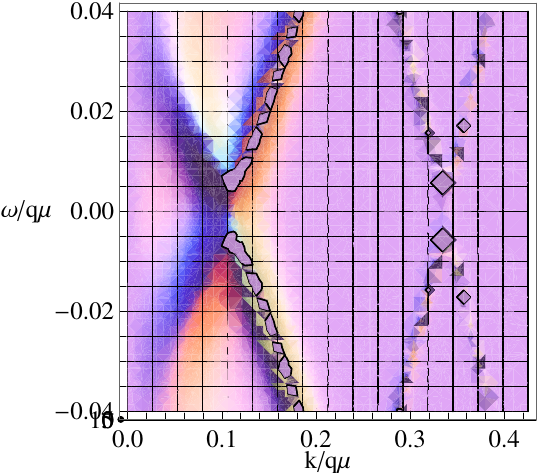}} 
\subfigure[$\theta=45^{\circ}$]{
\includegraphics[scale=0.75]{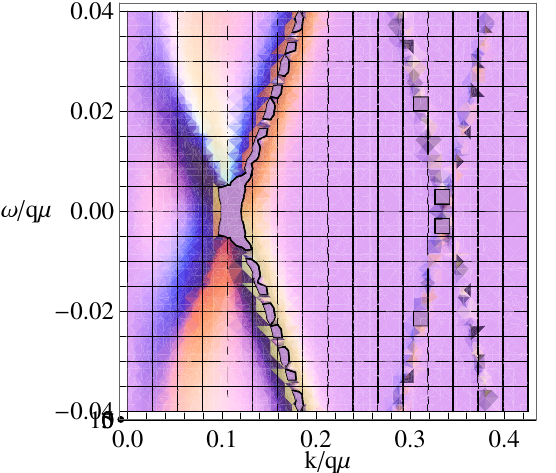}}
\end{center}
\caption{The spectral function $A(\protect\omega /q\protect\mu ,k/q\protect%
\mu )$ shown for various angles $\protect\theta $ with respect to the $x$%
-axis for $m^2=2, T=0.66T_{c}$ and $\protect\eta =0.15q$. }
\label{fig:angle}
\end{figure}
This case was first studied in Ref.\cite{Benini:2010qc}. It was found that
the spectral function at $\omega =0$ does not vanish in the nodal direction
but vanishes (up to some small values at small $k=\left\vert \mathbf{k}%
\right\vert $) in other directions. This suggests that the system is gapped
except in the nodal direction for $T<T_{1}$. At higher $T$, the ungapped
directions (the so called \textquotedblleft fermi arc\textquotedblright )
increase and eventually the system becomes ungapped above $T_{2}$ (with $%
T_{2}<T_{c}$). This is reflected in the growing angular size of fermi arc
with non-zero spectral function at $T_{1}<T<T_{2}$. In this region, a
typical spectral function $A(\omega /q\mu ,k/q\mu )$ for various angles $%
\theta $ with respect to the $x$-axis, is shown in Fig. \ref{fig:angle}. Up
to the small structure at low $k$, one can see the gap-like structure at $%
0^{\circ }$ and $30^{\circ }$ and the vanishing of the gap at $45^{\circ }$.
At $0^{\circ }$, the spectral width is significantly smaller than the gap,
suggesting it is a fermi liquid. One can also identify the fermi momentum at 
$k/q\mu \simeq 0.3$. In this model, the angular dependence of the fermi
momentum is not built in. It would be interesting to generalize this model
to incorporate this feature in the future.

At this point we want to note the appearance of another inner fermi arc in
the fermionic spectra function for a higher value of spin two field mass.
This is also clearly visible in Fig. \ref{fig:angle}. As we decrease the
mass of the background spin two field, this inner fermi surface disappears
but outer fermi arc still exist. It is interesting to study this in more
details and possibly ascribe any physical significance to this. Apparently
the existence of two fermi surfaces suggest that in the boundary field
theory, we may have two different characteristic fermionic degrees of
freedom.

\subsection{Dirac Mass Dependence}

While it is interesting to see the fermi arcs in this holographic model,
however, experimentally, the fermi arcs in high temperature SC's happen at $%
T_{1}=T_{c}<T<T_{2}$, i.e. fermi arcs happen in the pseudo gap phase but not
in the SC phase \cite{Norman:1998,Kanigel:2006,Kanigel:2007}.

With the goal of eliminating the fermi arc in the SC phase in mind, we study
the dirac mass ($m_{\zeta }$) effect to the spectral function. We found that
while $m_{\zeta }$ tends to increase the gap, it also makes the nodal points
not gapless any more.

\begin{figure}[tbp]
\begin{center}
\subfigure[$m_\zeta=0$]{
\includegraphics[scale=0.7]{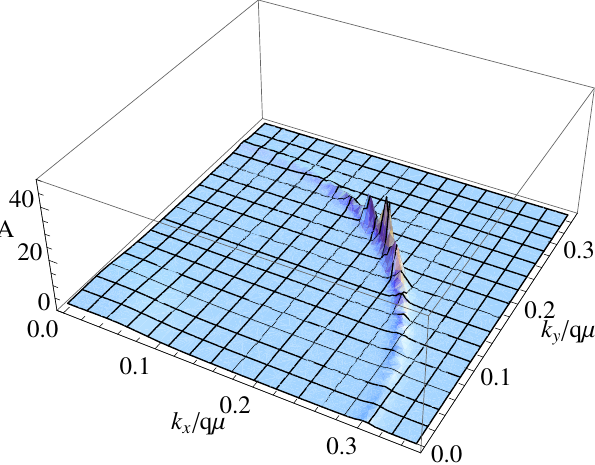}} 
\subfigure[$m_\zeta=0.2$]{
\includegraphics[scale=0.7]{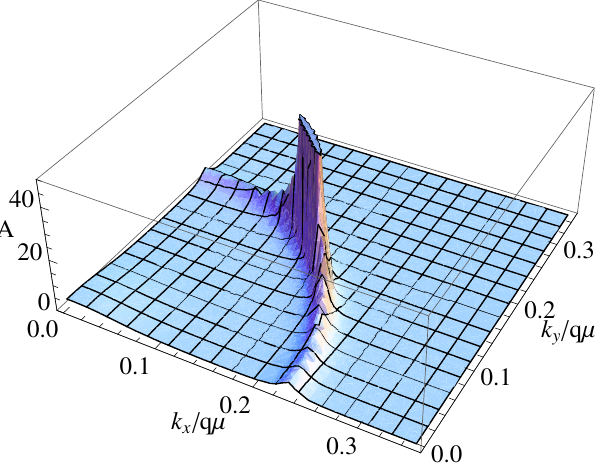}} 
\subfigure[$m_\zeta=0.4$]{
\includegraphics[scale=0.7]{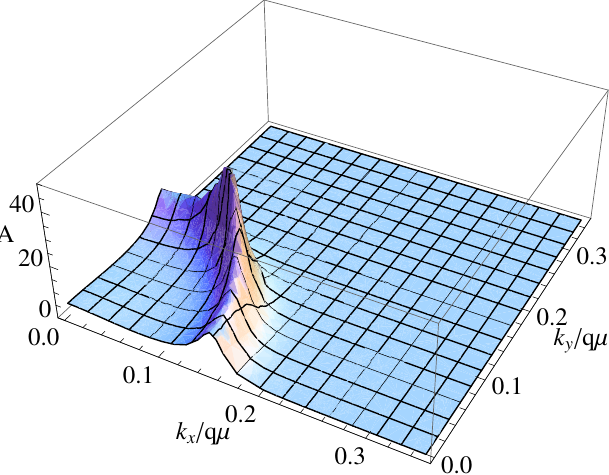}}
\end{center}
\caption{The dirac mass dependence of the spectral function $A(\protect%
\omega =0,\mathbf{k}/q\protect\mu )$. The fermi momentum decreases when $m_{%
\protect\zeta }$ is reduced. $m^2=0, T=0.66T_{c}$ and $\protect\eta =0.05q$.}
\label{fig:DiracMass}
\end{figure}

In Fig.\ref{fig:DiracMass}, it is shown that the fermi surface shrinks and
height of the peak is reduced when we turn on $m_{\zeta }$. If $m_{\zeta }$
is sufficiently large, the spectral function will be gapped in all
directions. 

\section{$d+id$ superconductors}

As mentioned above, if $\varphi _{xy}/\varphi _{xx}$ is not real, then we
cannot rotate the coordinates such that $\varphi _{xy}=0$, such that both $%
\varphi _{xx}$ and $\varphi _{xy}$ will exist. A particular interesting case
is $\varphi _{xy}/\varphi _{xx}=\pm i$, whose ratio is unchanged under the
rotation in Eq. (\ref{rotation}). This means the difference between $\varphi
_{xx}$ and $\varphi _{xy}$ in all directions is just a common phase which
can be gauged away without physical consequences. This implies the fermion
spectral function will be the same in all directions.

\begin{figure}[tbp]
\begin{center}
\subfigure[$\gamma=\infty$]{
\includegraphics[scale=0.7]{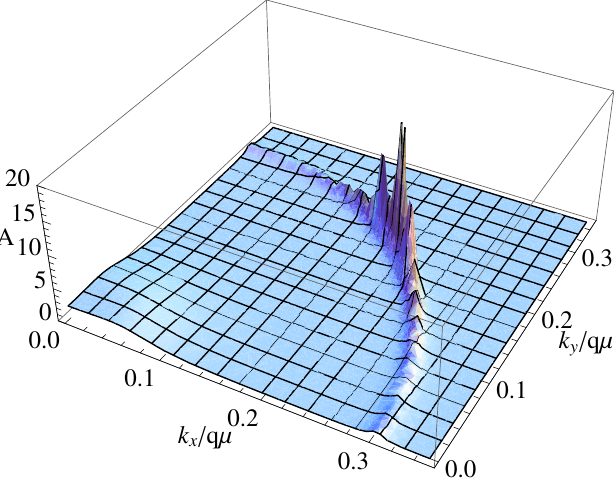}} 
\subfigure[$\gamma=2$]{
\includegraphics[scale=0.7]{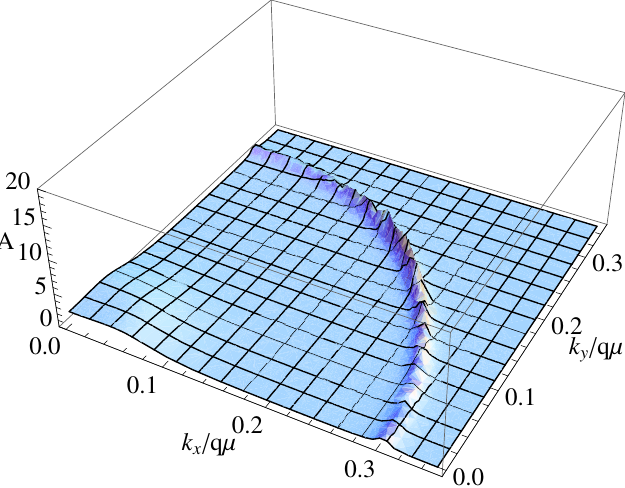}} 
\subfigure[$\gamma=1$]{
\includegraphics[scale=0.7]{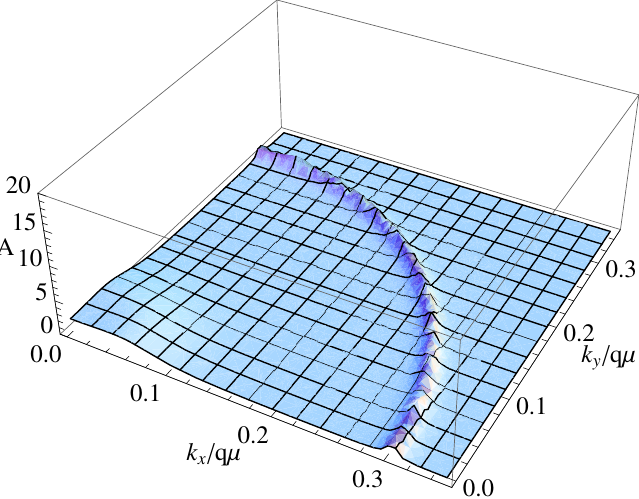}} 
\subfigure[$\gamma=\frac{1}{2}$]{
\includegraphics[scale=0.7]{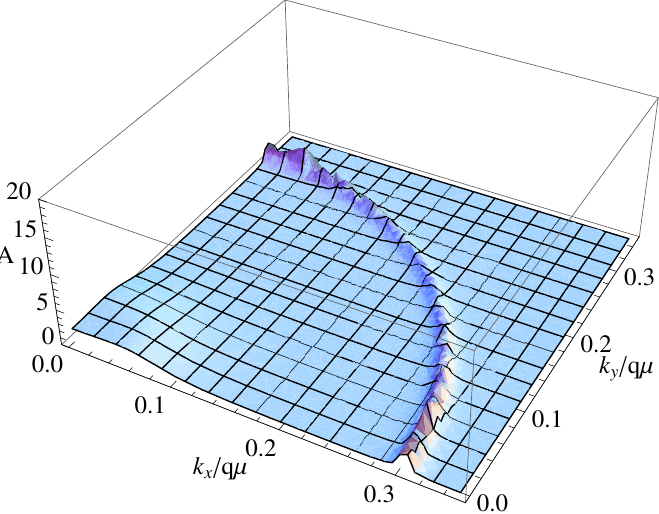}} 
\subfigure[$\gamma=0$]{
\includegraphics[scale=0.7]{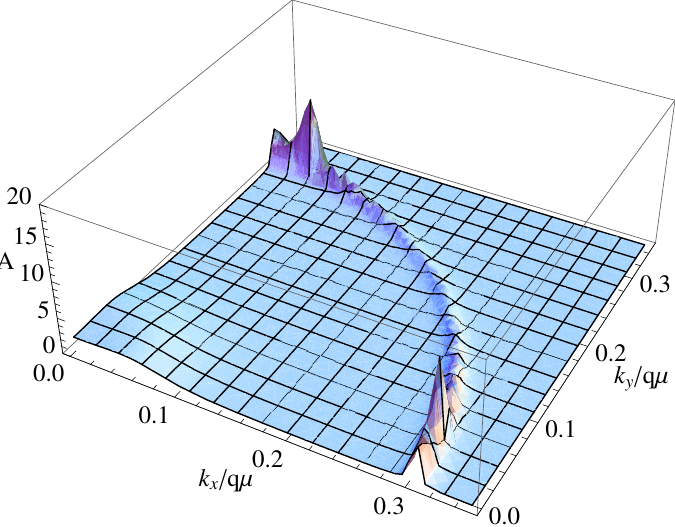}}
\end{center}
\caption{$A(\protect\omega =0,\mathbf{k}/q\protect\mu )$ for $d+id$
superconductors. For $\protect\gamma =\infty $ and $\protect\gamma =0$, the
spectral functions reduce to the d-wave case. For, $\protect\gamma =1$, the
spectral function is isotropic and is s-wave like. $m^2=0, T=0.66T_{c}$, $%
\protect\eta =0.05q$, $m_{\protect\zeta }=0$}
\label{fig:d+id}
\end{figure}

In Fig.\ref{fig:d+id}, we plot $A(\omega =0,\mathbf{k})$ for different 
\begin{equation}
\gamma \equiv \frac{i\varphi _{xy}}{\varphi _{xx}}\text{.}
\end{equation}%
As expected, the $\gamma =0$ case is just a $\pi /4$ rotation of the $\gamma
\rightarrow \infty $ case. Both of them are purely d-wave SC's. In the $d+id$
SC with $\gamma =1$, the spectral function looks like a s-wave SC.

\section{Conductivity}

\label{sec5}

In this section we will calculate the conductivity by turning on an
electromagnetic perturbation as previously discussed. According to the
standard AdS/CFT dictionary, the gauge field perturbation in the bulk will
lead to a boundary current. The boundary value of the perturbed gauge field
becomes a source for this boundary current.

The conductivity tensor $\sigma _{ij}$ can be defined as 
\begin{equation}
J_{i}=\sigma _{ij}E_{j},  \label{condeq}
\end{equation}%
where $E_{i}$ and $J_{i}$ are the external electric field and the induced
current, respectively, in the $i$-direction ($i=x,y$). In linear response
theory, $\sigma _{ij}$\ is a current-current correlator which can be
schematically denoted as $\sigma _{ij}\sim \left\langle \Omega \left\vert %
\left[ J_{i},J_{j}\right] \right\vert \Omega \right\rangle $, where the
matrix element denotes an ensemble average. Under a $\pi /2$\ rotation along
the $z$-axis ($R$), $R^{-1}J_{i}R=\epsilon _{ij}J_{j}$, where $\epsilon
_{ij} $\ is an anti-symmetric tensor, and assuming the ensemble average is
governed by properties of the ground state which is in general a d+id
condensation, we have $R\left\vert \Omega \right\rangle =-\left\vert \Omega
\right\rangle $. Then, $\sigma _{ij}\sim $\ $\left\langle \Omega \left\vert
RR^{-1}\left[ J_{i},J_{j}\right] RR^{-1}\right\vert \Omega \right\rangle
=\left\langle \Omega \left\vert R^{-1}\left[ J_{i},J_{j}\right] R\right\vert
\Omega \right\rangle =\left\langle \Omega \left\vert \left[ \epsilon
_{ik}J_{k},\epsilon _{jl}J_{l}\right] \right\vert \Omega \right\rangle $.
This implies $\sigma _{xx}=\sigma _{yy}$\ and $\sigma _{xy}=-\sigma _{yx}$.

Under a parity ($P_{x}$) transformation with respect to the $x$-axis, the
condensates transform as 
\begin{equation}
P_{x}\left( 
\begin{array}{cc}
\langle \mathcal{O}_{1}\rangle & \langle \mathcal{O}_{2}\rangle \\ 
\langle \mathcal{O}_{2}\rangle & -\langle \mathcal{O}_{1}\rangle%
\end{array}%
\right) \rightarrow \left( 
\begin{array}{cc}
\langle \mathcal{O}_{1}\rangle & -\langle \mathcal{O}_{2}\rangle \\ 
-\langle \mathcal{O}_{2}\rangle & -\langle \mathcal{O}_{1}\rangle%
\end{array}%
\right) .
\end{equation}%
If $\langle \mathcal{O}_{2}\rangle =0$, as the case which is always
achievable in a d-wave SC, then $P_{x}\left\vert \Omega \right\rangle =\pm
\left\vert \Omega \right\rangle $. Thus, $\sigma _{ij}\sim \left\langle
\Omega \left\vert P_{x}P_{x}^{-1}\left[ J_{i},J_{j}\right]
P_{x}P_{x}^{-1}\right\vert \Omega \right\rangle =\left\langle \Omega
\left\vert P_{x}^{-1}\left[ J_{i},J_{j}\right] P_{x}\right\vert \Omega
\right\rangle \sim (-1)^{\delta _{iy}+\delta _{jy}}\sigma _{ij}$. This
yields $\sigma _{xy}=\sigma _{yx}=0$. In the case of $d+id$, however, $%
\left\vert \Omega \right\rangle $ is not an eigenstate of $P_{x}$. Then, in
general $\sigma _{xy}=-\sigma _{yx}\neq 0$. A similar argument using time
reversal symmetry yields the same conclusion. So unlike the normal
holographic d-wave SC, in addition to standard $\sigma _{xx}$, the $d+id$ SC
could have a non-vanishing Hall conductivity $\sigma _{xy}$ component.

In order to make our analysis simple, using the above symmetry we can
diagonalize our conductivity tensor such that Eq.\ref{condeq} becomes 
\begin{equation}
J_{\pm }=\sigma _{\mp }E_{\pm },
\end{equation}%
where 
\begin{equation}
~J_{\pm }=J_{x}\pm iJ_{y}~;~E_{\pm }=E_{x}\pm iE_{y}~;~\sigma _{\pm }=\sigma
_{xx}\pm i\sigma _{xy}.
\end{equation}

As we have mentioned before, in order to calculate the conductivity, we need
to consider the linearized perturbations in the bulk black hole background 
\cite{son}. A consistent set of gauge field and spin-2 field perturbations
are 
\begin{eqnarray}
\delta A(t,z) &=&e^{-i\omega t}(a_{x}(z)dx+a_{y}(z)dy),~~~~~  \notag \\
\delta \varphi _{\mu \nu }(t,z) &=&e^{-i\omega t}\left( 
\begin{array}{cccc}
0 & 0 & \xi _{tx}(z) & \xi _{ty}(z) \\ 
0 & 0 & \xi _{rx}(z) & \xi _{ry}(z) \\ 
\xi _{tx}(z) & \xi _{rx}(z) & 0 & 0 \\ 
\xi _{ty}(z) & \xi _{ry}(z) & 0 & 0%
\end{array}%
\right) ,
\end{eqnarray}%
where $\omega $ is the frequency of the perturbation. Similarly, we can
define the perturbation for the complex conjugate field $\delta \varphi
_{\mu \nu }^{\ast }$ also. In the above perturbation ansatz we have only
considered zero momentum modes. We only limit ourself to consider the case
where $\gamma =i\varphi _{xy}/\varphi _{xx}=Q_{2}/Q_{1}$ is real.

With this perturbation ansatz we would like to obtain the linear order
perturbation equations. Apparently as one can observe from \cite%
{Benini:2010pr} that when the background has only d-wave condensation, the
equation of motion for two independent gauge field perturbations $a_{x}$ and 
$a_{y}$ are not mixed each other. From this one can arrive at the conclusion
that there is no Hall current in the dual superfluid phase. But in a $d+id$
SC, $a_{x}$ and $a_{y}$ couple which leads to Hall conductivity. The
equation of motion in this general background is complicated to solve. But
two decoupled set of equations emerge if we perform the following field
redefinition: 
\begin{eqnarray}
a_{m} &=&a_{x}+ia_{y},~a_{p}=a_{x}-ia_{y},~~  \notag \\
\xi _{mt} &=&\xi _{ty}-i\xi _{tx},\ \xi _{pt}=\xi _{ty}+i\xi _{tx},  \notag
\\
\xi _{mz} &=&\xi _{zy}-i\xi _{zx},~\xi _{pz}=\xi _{zy}+i\xi _{zx}{,}  \notag
\\
\xi _{mt}^{\ast } &=&\xi _{ty}^{\ast }-i\xi _{tx}^{\ast },~\xi _{pt}^{\ast
}=\xi _{ty}^{\ast }+i\xi _{tx}^{\ast },~  \notag \\
\xi _{mz}^{\ast } &=&\xi _{zy}^{\ast }-i\xi _{zx}^{\ast },~\xi _{pz}^{\ast
}=\xi _{zy}^{\ast }+i\xi _{zx}^{\ast }.
\end{eqnarray}%
Under this field redefinition one set of decoupled equations of motion
become 
\begin{eqnarray}
a_{m}^{\prime \prime }+\frac{f^{\prime }}{f}a_{m}^{\prime } &+&\frac{\omega
^{2}}{f^{2}}a_{m}+\frac{q\psi }{2f^{2}}\left[ (Q_{2}-Q_{1})\xi _{mt}^{\ast
}-(Q_{2}+Q_{1})\xi _{mt}\right] -i\frac{q\psi }{2}\left[ (Q_{2}-Q_{1})\xi
_{mz}^{\ast ^{\prime }}\right.  \notag \\
&-&\left. (Q_{2}+Q_{1})\xi _{mz}^{\prime }\right] +i\frac{q}{2f}(\psi
^{\prime }f-f^{\prime }\psi )\left[ (Q_{2}-Q_{1})\xi _{mz}^{\ast
}-(Q_{2}+Q_{1})\xi _{mz}\right] =0,  \notag \\
\xi _{mt}^{\prime \prime }+\frac{2}{z}\xi _{mt}^{\prime } &-&\frac{%
2f+L^{2}m^{2}}{z^{2}f}\xi _{mt}+\frac{L^{2}q\psi }{4z^{2}f}%
(Q_{2}+Q_{1})(\omega +2q\phi )a_{m}  \notag \\
&&\quad \quad \quad \quad ~~~~~~~~~~~~~~~~~~~~+\frac{i}{2}\left[ (2(\omega
+q\phi )\xi _{mz}^{\prime }+q\phi ^{\prime }\xi _{mz}\right] =0,  \notag \\
\xi _{mz}\left[ z^{2}(\omega \right. &+&\left. q\phi )^{2}-m^{2}L^{2}f\right]
+i\frac{L^{2}qf}{4}(Q_{1}+Q_{2})(\psi a_{m}^{\prime }+2\psi ^{\prime }a_{m})
\notag \\
&&~~~~~~~~~~~~~~-iz^{2}(\omega +q\phi )\xi _{mt}^{\prime }-i\frac{z}{2}%
(4\omega +4q\phi +qz\phi ^{\prime })\xi _{mt}=0,  \notag \\
\xi _{mt}^{\ast ^{\prime \prime }}+\frac{2}{z}\xi _{mt}^{\ast ^{\prime }} &-&%
\frac{2f+L^{2}m^{2}}{z^{2}f}\xi _{mt}^{\ast }+\frac{L^{2}q\psi }{4z^{2}f}%
(Q_{2}-Q_{1})(-\omega +2q\phi )a_{m}  \notag \\
&&\quad \quad \quad ~~~~~~~~~~~~~~~~~~~~-\frac{i}{2}\left[ (2(-\omega +q\phi
)\xi _{mz}^{\ast ^{\prime }}+q\phi ^{\prime }\xi _{mz}^{\ast }\right] =0, 
\notag \\
\xi _{mz}^{\ast }\left[ z^{2}(-\omega \right. &+&\left. q\phi
)^{2}-m^{2}L^{2}f\right] -i\frac{L^{2}qf}{4}(Q_{2}-Q_{1})(\psi a_{m}^{\prime
}+2\psi ^{\prime }a_{m})  \notag \\
&&~~~~~~~~~+iz^{2}(-\omega +q\phi )\xi _{mt}^{\ast ^{\prime }}+i\frac{z}{2}%
(-4\omega +4q\phi +qz\phi ^{\prime })\xi _{mt}^{\ast }=0,  \label{coupled}
\end{eqnarray}%
As we can see from the above set of equations, two equations for $\xi _{mz}$
and $\xi _{mz}^{\ast }$ are algebraic in nature so we can use the third and
the fifth equations to substitute $\xi _{mz}$ and $\xi _{mz}^{\ast }$ in
other equations. Therefore, we have only three coupled fields equations for $%
a_{m},\xi _{mt}$ and $\xi _{mt}^{\ast }$ to be solved. Similarly we have
another set of equations for $a_{p},\xi _{pt},\xi _{pt}^{\ast },\xi _{pz}$
and $\xi _{pz}^{\ast }$ components. These equations can be obtained simply
by replacing $m\rightarrow p$ and $Q_{1}\rightarrow -Q_{1}$ in the above.

Now we will solve the above set of equations numerically by integrating them
from the horizon to the boundary of the bulk AdS black hole spacetime. There
exists a well defined procedure to calculate the retarded Greens function of
the dual boundary field theory \cite{son,Iqbal:2009fd}. At the horizon, the
ingoing wave boundary condition of all the fields are imposed to ensure
causality: 
\begin{eqnarray}
a_{m} &=&f^{-i\frac{\omega }{3}}~a_{0}+\cdots  \notag \\
\xi _{mt} &=&f^{-i\frac{\omega }{3}}~\xi _{mt0}+\cdots ~~;~~\xi _{mt}^{\ast
}=f^{-i\frac{\omega }{3}}~\xi _{mt0}^{\ast }+\cdots  \notag \\
\xi _{mz} &=&f^{-i\frac{\omega }{3}-1}~\xi _{mz0}+\cdots ~~;~~\xi
_{mz}^{\ast }=f^{-i\frac{\omega }{3}-1}~\xi _{mz0}^{\ast }+\cdots
\end{eqnarray}%
Where $a_{0},\xi _{mt0}$ and $\xi _{mt0}^{\ast }$ are some arbitrary complex
constants yet to be determined. Since the system in Eq. \ref{coupled} is
linear and homogeneous, we can set $a_{0}=1$ because this only affects the
overall normalization of $\xi _{mt0}$ and $\xi _{mt0}^{\ast }$. It has no
effect on conductivity. So by demanding $\xi _{mt}$ and $\xi _{mt}^{\ast }$
to be normalizable near the boundary as was done in Ref. \cite{Benini:2010pr}%
, $\xi _{mt0}$ and $\xi _{mt0}^{\ast }$ are uniquely fixed. Therefore we can
solve for the conductivity. This can all be done without the shooting
method. Because the system is linear and homogeneous, the problem is reduced
to taking the correct linear combination of three independent solutions to
eliminate the non-normalizable terms of $\xi _{mt}$ and $\xi _{mt}^{\ast }$.
Once this is done, we can use the standard AdS/CFT technique to extract the
retarded Greens function from the solution of the electromagnetic
perturbation $a_{m}.$

\begin{figure}[tbp]
\begin{center}
\includegraphics[scale=1.0]{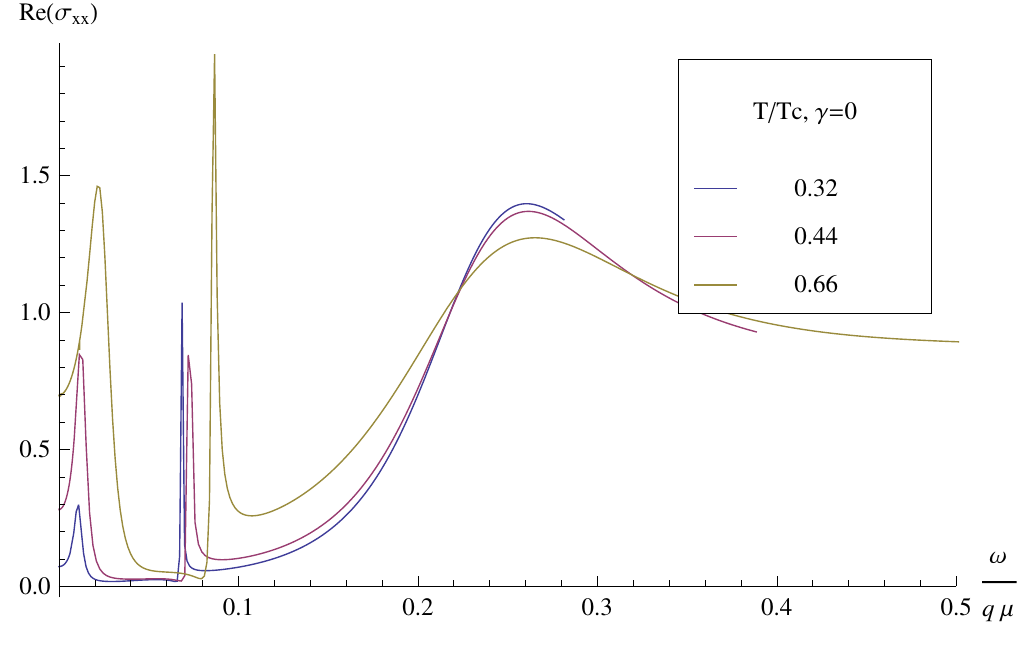}
\end{center}
\caption{The real part of the normal conductivity $Re[\protect\sigma _{xx}]$
vs. $T/T_{c}$ for holographic d-wave superconductors ($\protect\gamma =0$)
with $m^{2}=4$. There are $\protect\delta (\protect\omega )$ type
supercurrent contributions in these curves that cannot be seen clearly. The
hall conductivity $\protect\sigma _{xy}$ vanishes. }
\label{Q0sigxx}
\end{figure}

\begin{figure}[tbp]
\begin{center}
\includegraphics[scale=1.0]{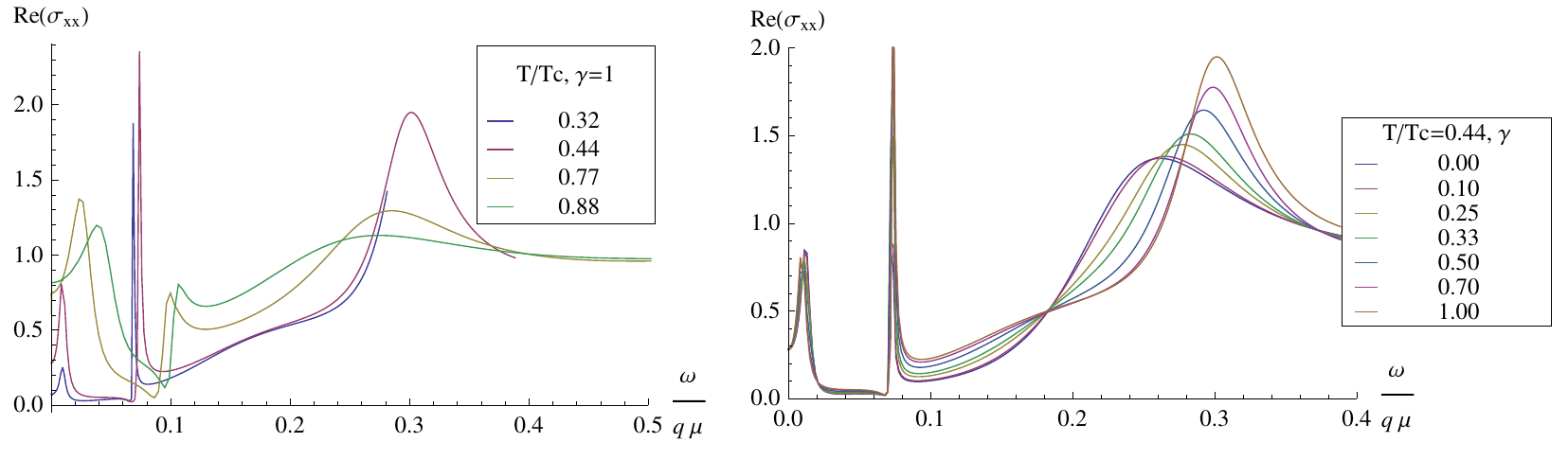}%Q_1sigxx.pdf}
\end{center}
\caption{$Re[\protect\sigma _{xx}]$ of
holographic $d+id$ superconductors ($m^{2}=4$) for $\protect\gamma =1$ with
various $T$ (left panel) and for $T/T_{c}=0.44$ with various $\protect\gamma 
$ (right panel). There are $\protect\delta (\protect\omega )$ type
supercurrent contributions in these curves that cannot be seen clearly.}
\label{Q1sigxx}
\end{figure}

\begin{figure}[tbp]
\begin{center}
\includegraphics[scale=1.0]{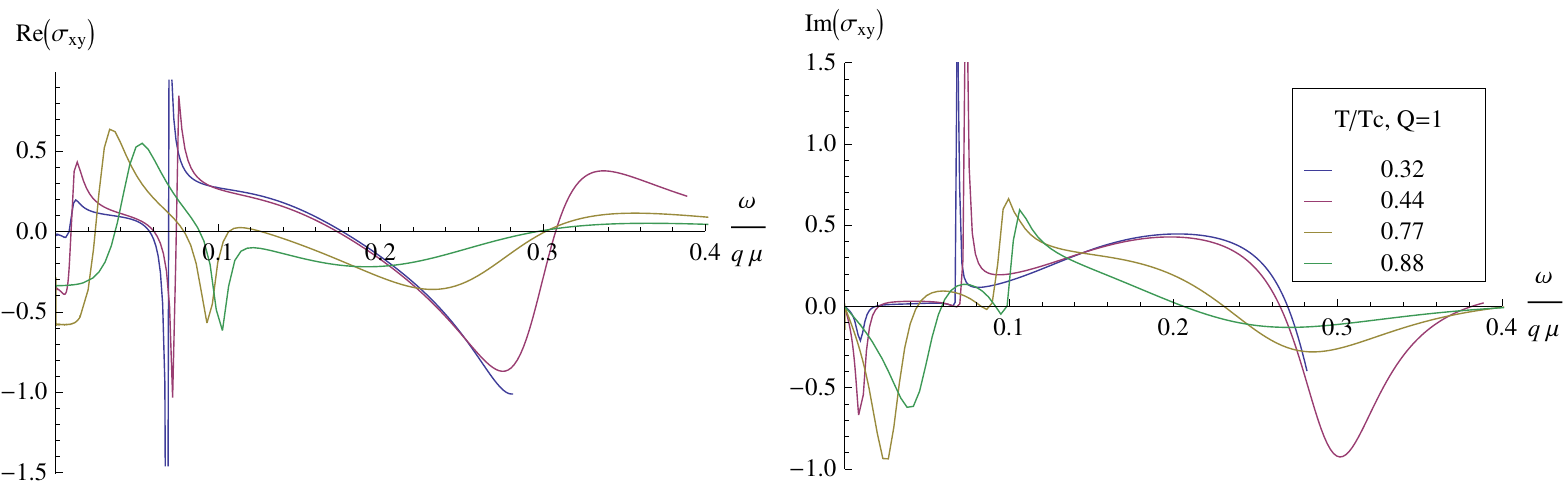}
\end{center}
\caption{The real and imaginary parts of the Hall conductivity, $Re[\protect%
\sigma _{xy}]$ and $Im[\protect\sigma _{xy}]$, for $d+id$ superconductors
with $m^{2}=4$, $\protect\gamma =1$ and various $T$.}
\label{Q1sigxy}
\end{figure}

Near the boundary, $a_{m}$ has the asymptotic behavior 
\begin{equation}
a_{m}(z)=a_{m}^{0}+a_{m}^{1}z+\cdots ,
\end{equation}%
where $a_{m}^{0}$ corresponds to an applied electric field in the boundary
theory and $a_{m}^{1}$ corresponds to the induced current. The standard
definition of the Greens function \cite{son,Iqbal:2009fd} is 
\begin{equation}
\sigma _{+}=\sigma _{xx}+i\sigma _{xy}=\lim_{z\rightarrow 0}\frac{-ia_{m}^{1}%
}{\omega a_{m}^{0}}.
\end{equation}%
Then the standard conductivity tensor component would simply be 
\begin{equation}
\sigma _{xx}=\frac{1}{2}(\sigma _{+}+\sigma _{-}),~\sigma _{xy}=\frac{-i}{2}%
(\sigma _{+}-\sigma _{-}).  \label{sigma}
\end{equation}

\begin{figure}[tbp]
\begin{center}
\includegraphics[scale=1.0]{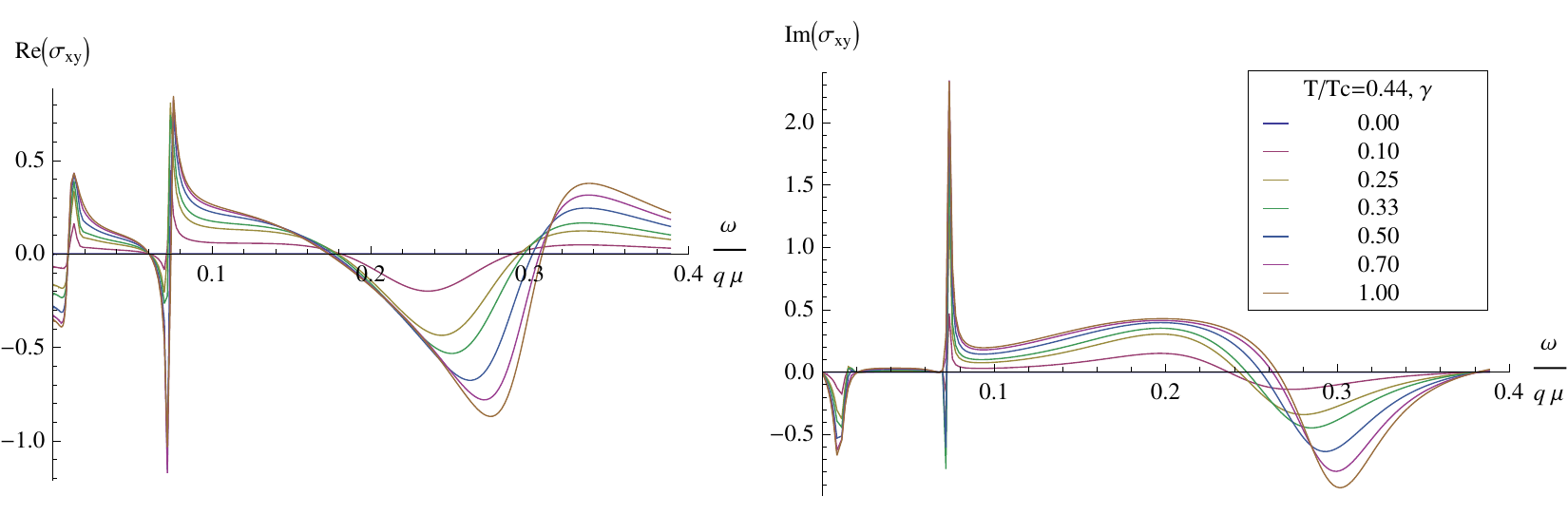}%Q_1sigxx.pdf}
\end{center}
\caption{
$Re[\protect\sigma _{xy}]$ and $%
Im[\protect\sigma _{xy}]$ for various $\protect\gamma $ at $m^{2}=4$ and  $%
T/T_{c}=0.44$. $T_{c}$ is independent of $\protect\gamma $. $\protect\sigma %
_{xy}=0$ when $\protect\gamma =0$.
}
\label{gammasigxy}
\end{figure}

In the d-wave case with $\varphi _{xy}=\gamma =0$, $\sigma _{xy}=0$, and we
have reproduced the $\sigma _{xx}$ result of Refs.\cite{Benini:2010qc,
Benini:2010pr} in Fig.\ref{Q0sigxx}. For $\gamma \neq 0$, we have only shown
the $0\leq \gamma \leq 1$ result in Figs.\ref{Q1sigxx}-\ref{gammasigxy}
because $\gamma \rightarrow 1/\gamma $ under a $\pi /2$ rotation and then 
\begin{equation}
\sigma _{ij}\left( \gamma \right) =\sigma _{ij}\left( 1/\gamma \right) .
\end{equation}%
In each case, the real part of $\sigma _{xx}$, Re($\sigma _{xx}$), has a
delta function type contribution $\delta (\omega )$ from the super current
if the system is in the superconducting phase. Also, Re($\sigma _{xx}$) is
always non-negative. Below some frequency $\omega _{s}$, Re($\sigma _{xx}$)
could have several peaks, suggesting there are several spin one resonance
states with masses set by the scale $q\mu $. Above $\omega _{s}$, Re($\sigma
_{xx}$) reaches its asymptotic value which is one. Numerically, $\omega
_{s}\sim q\mu $. Unlike the s-wave SC's, the fermi surfaces are not gapped
everywhere for the d-wave or $d+id$ SC's. These conducting electrons can
respond to an electricity field of any frequency. Thus, there should be no
gap in the conductivity. Our result also has this feature. 

For Re($\sigma _{xy}$), there is no $\delta (\omega )$ contribution and it
does not have a definite sign. At the peaks of Re($\sigma _{xx}$), Re($%
\sigma _{xy}$) vanishes (see Figs. \ref{Q1sigxx} and \ref{Q1sigxy}). (But Re(%
$\sigma _{xy}$) vanishes does not imply a peak at Re($\sigma _{xx}$).) This
is because suppose one of the diagonalized conductivity $\sigma _{+}$ is
much bigger than the other in size, then according to Eq. (\ref{sigma}), Re($%
\sigma _{xy}$) is nothing but Im($\sigma _{xx}$). So it relates to Re($%
\sigma _{xx}$) as described above. 

In Figs. \ref{Q1sigxy} and \ref{gammasigxy}, the $\sigma _{xy}$ for various $T
$ and $\gamma $ is shown. It is interesting that within a small frequency window,
 the Hall conductivity Re($%
\sigma _{xy}$) could be very sensitive to the external
electric field frequency with the Hall conductivity changes dramatically from a large negative
value to a large positive value. Thus, it could
be a good electromagnetic wave frequency sensor. About the $\gamma $
dependence, there is a nearly universal Re($\sigma _{xx}$) in Fig. \ref%
{Q1sigxx} when $\omega /q\mu \simeq 0.18$. Also, the shape of $\sigma _{xy}$
in Fig. \ref{gammasigxy} is insensitive to $\gamma $. Those are curious features of this
type of superconductors.

\section{Conclusion}

We studied a holographic model of $d+id$ superconductors based on the action
proposed by Benini, Herzog, and Yarom \cite{Benini:2010qc}. The model
contains a charged spin two field in an AdS black hole spacetime. Working in
the probe limit, the normalizable solution of the spin two field in the bulk
gives rise to a $d+id$ superconducting order parameter at the boundary of
the AdS.

We have calculated the fermion spectral function in this superconducting
background and confirmed the existence of fermi arcs for non-vanishing
Majorana couplings. Depending on the relative strength $\gamma $ of the $d$
and $id$ condensations, we found that the position and the size of the fermi
arcs are changing. Specifically when we take $\gamma =1$, the spectral
functions become isotropic and are s-wave like. We also studied fermion mass
effect. By changing the fermion mass, we saw the fermi momentum is changing.
We have also calculated the conductivity for these holographic $d+id$
superconductors where time reversal symmetry has been broken spontaneously.
A non-vanishing Hall conductivity has been obtained even without an external
magnetic field.

As we know in a real high temperature superconductor, the fermi arc has been
observed in a pseudo gap phase before the superconducting phase transition.
So far in the existing holographic models of d-wave superconductor the fermi
arc appears in the superconducting phase. This is an interesting open
problem to construct such a holographic model.

\begin{acknowledgments}
We thank Chyh-Hong Chern, Ying-Jer Kao, Feng-Li Lin, Wen-Yu Wen for useful
discussions and Brian Smigielski for a careful reading of the manuscript.
This work was supported in part by the LeCosPa, NSC and NCTS of R.O.C..
\end{acknowledgments}

%%%%%%%%%%%%%%%%%%%%%%%%%%%%%%%%%%%%%%%%%%%%%%%%%%%%%%%%%%%%%%

\end{document}